\documentclass{aa}
\usepackage{graphicx}
\usepackage{natbib}
\usepackage[varg]{txfonts}

\newcommand\XMM{XMM-\emph{Newton}}
\newcommand\xmmn{XMM-\emph{Newton}}
\newcommand\Swift{\emph{Swift}}
\newcommand\swi{\emph{Swift}}
\newcommand\cxo{\emph{Chandra}}
\newcommand\hst{\emph{Hubble}}
\newcommand\GALEX{\emph{GALEX}}

\newcommand\ovb{OV~Bootis}
\newcommand\ovbs{OV~Boo\xspace}
\newcommand\jt{SDSSJ1035\xspace}
\newcommand\cgs{erg\,cm$^{-2}$\,s$^{-1}$}
\newcommand\fcgs{erg\,cm$^{-2}$\,s$^{-1}$\,\AA$^{-1}$}
\newcommand\msun{M$_\odot$}
\newcommand\rsun{R$_\odot$}
\newcommand\rat{s$^{-1}$}
\newcommand\yrat{yr$^{-1}$}
\newcommand\teff{$T_{\rm eff}$}

\begin{document}

\title{X-ray and ultraviolet observations of the eclipsing cataclysmic variables \ovb\ and SDSS\,J103533.02+055158.3 with degenerate donors\thanks{Based on observations obtained with \xmmn, an ESA science mission with instruments and contributions directly funded by ESA Member States and NASA}}

\author{Axel Schwope\inst{1}
\and Hauke Worpel\inst{1}
\and Iris Traulsen\inst{1}
}
\institute{Leibniz-Institut f\"ur Astrophysik Potsdam (AIP), An der Sternwarte 16, 14482 Potsdam, Germany}

\authorrunning{Schwope et al.}
\titlerunning{The eclipsing cataclysmic variables \ovb\ and SDSS1035+05}
\date{\today}


\abstract
{The majority of cataclysmic variables are predicted to be post-period minimum systems with degenerate donor stars, the period bouncers. Owing to their intrinsic faintness, however, only a handful of these systems have so far been securely identified.}
{We want to study the X-ray properties of two eclipsing period bouncers, OV Bootis and SDSS J103533.02+055158.3, that were selected for this study due to their proximity to Earth.}
{We have obtained \XMM\ phase-resolved X-ray and ultraviolet observations of the two objects for spectral and timing analysis.}
{Owing to a recent dwarf nova outburst \ovb\ was much brighter than SDSS J103533.02+055158.3 at X-ray and ultraviolet wavelengths and the eclipse could be studied in some detail. An updated eclipse ephemeris was derived. The X-rays were shown to originate close to the white dwarf, the boundary layer, with significant absorption affecting its spectrum. There was no absorption in SDSS J103533.02+055158.3, despite being observed at the same inclination indicating different shapes of the disk and the disk rim. The white-dwarf temperature was re-determined for both objects: the white dwarf in \ovb\ was still hot (23,000\,K) five months after a dwarf nova outburst, and the white dwarf in SDSS J103533.02+055158.3 hotter than assumed previously (\teff $= 11,500$\,K).}
{All three cataclysmic variables with degenerate donors studied so far in X-rays, including SDSS \,J121209.31+013627.7, were clearly discovered in X-rays and revealed mass accretion rates $\dot{M} \geq 8\times 10^{-15}$\,\msun\,\yrat. 
If their X-ray behavior is representative of the subpopulation of period bouncers, the all-sky X-ray surveys with eROSITA together with comprehensive follow-up will uncover new objects in sufficient number to address the remaining questions concerning late-stage cataclysmic variable evolution.}

\maketitle

\section{Introduction}

Cataclysmic variables (CVs) consist of an accreting white dwarf (WD) primary and a low-mass Roche-lobe filling secondary. These systems represent the most common endpoint of close binary evolution. Below the period gap, gravitational radiation drives the period of the binary towards shorter periods. Eventually, the donor star crosses the hydrogen burning limit, at which point further mass loss causes the donor to expand. This, in turn, leads to an increase in the orbital period, and these systems are therefore referred to as {period bouncers}.

Although around 70\% of CV systems are predicted to be post-period minimum \citep{goliasch_nelson15, belloni+20}, examples of this class have been difficult to identify owing to their intrinsic faintness. Furthermore, to distinguish between pre- and post-period minimum systems it is not enough to show that the system has a low accretion rate. The mass of the secondary must be determined to show that the donor mass is now too small to be burning hydrogen. This can be done by performing optical spectroscopy  (e.g., \citealt{SchwopeChristensen2010}), or radius measurements in eclipsing systems \citep{BurleighEtAl2006, LittlefairEtAl2008}. It is likely that many of the secondaries in period bouncer systems are substellar in mass \citep{SavouryEtAl2011}.

To date, only a handful of unimpeachable members of the class have been identified, which were studied intensively in the optical, but little is known about their X-ray emission as a sign of accretion. 
Here we present an X-ray study of two confirmed period bouncers: \ovb\ and SDSS~J103533.02+055158.4 (hereafter \ovbs and \jt, respectively). These two systems were predicted to be the most X-ray bright of the known period bouncers simply due to their vicinity \citep[$160\pm10$\,pc for \ovbs and $171\pm10$\,pc for \jt;][respectively]{LittlefairEtAl2007,LittlefairEtAl2008} and were proposed to be observed by \XMM. 

OV Boo was identified as a CV from Sloan Digital Sky Survey (SDSS) spectra, and follow-up observations showed it to be a deeply eclipsing system with a 1.12\,hr period \citep{SzkodyEtAl2005}. The period was refined to $66.61194\pm 0.00004$ minutes through intensive photometry by \cite{LittlefairEtAl2007} and \cite{PattersonEtAl2008}. The orbital period of \ovbs lies below the period minimum predicted by standard CV evolution, and could be explained if the system were formed directly from a detached WD--brown dwarf pair. The detailed observational studies performed by \cite{PattersonEtAl2008} and \cite{LittlefairEtAl2007,LittlefairEtAl2008} revealed a complex eclipse structure, which was resolved into the WD, the accretion disk, a bright spot on the rim of the disk, and the donor star. \cite{LittlefairEtAl2007} also derived parameters of the WD; they give a radius of $0.0091\pm0.0001$\,\rsun\ and a mass of $0.090\pm0.001$\,\msun, which we use in this paper.

From its high estimated proper motion of $\sim 160$\,mas\,yr$^{-1}$ it is believed to be a member of the Galactic halo rather than the disk \citep{PattersonEtAl2008}. This interpretation is supported by ultraviolet (UV) observations \citep{UthasEtAl2011} that showed it to have significantly subsolar metallicity. In 2017 \ovbs underwent a dwarf nova (DN) outburst, studied intensively \citep{PattersonEtAl2017, TanabeEtAl2018}, and may have had previous outbursts in 1906 and 1984 \citep{Bengtsson2017a, Bengtsson2017b}. The second Gaia data release gives \ovbs  a distance of $210^{+7}_{-6}$\,pc \citep{Gaia2018, Bailer-JonesEtAl2018} and confirms a high proper motion of the object of 155\,mas\,yr$^{-1}$.


\jt has a period of 82.10 minutes, and the donor mass is $M=0.052\pm 0.002 M_\odot$ \citep{LittlefairEtAl2006}. Its distance is $209^{+16}_{-14}$\,pc \citep{Gaia2016, Gaia2018, Bailer-JonesEtAl2018}. The optical spectrum is dominated by the WD. The companion, fainter than L0, is completely undetectable \citep{SouthworthEtAl2006}. Owing to the faintness of the source, little else is known about it.

We note that the Gaia distances of both objects, which we  use in this paper, are larger than originally estimated.

\section{Observations and methods}

\begin{table*}
 \caption{X-ray observation log}
 \begin{tabular}{llllll}
  Target    & Inst.    & Date         & ObsID       & Exp (ks) & Camera-Mode-Filter      \\
  \hline 
  OV Boo    & \cxo  & 2003-Mar-14  & 4071        & 4.9           \\
  OV Boo    & \Swift    & 2017-Mar-16  & 00035443004 & 1.0      & XRT-PC     \\
            &          &              &             &          & UVOT-IM-UVW1     \\
  OV Boo    & \Swift    & 2017-Mar-18  & 00035443005 & 1.0      & XRT-PC     \\
            &          &              &             &          & UVOT-IM-UVW1     \\
  OV Boo    & \Swift    & 2017-Mar-19  & 00035443006 & 1.0      & XRT-PC, XRT-WT   \\
            &          &              &             &          & UVOT-IM-UVW1     \\
  OV Boo    & \Swift    & 2017-Mar-21  & 00035443007 & 0.6      & XRT-PC           \\
            &          &              &             &          & UVOT-IM-UVW2, UVOT-EV-UVW2     \\
  OV Boo    & \Swift    & 2017-Mar-23  & 00035443008 & 0.9      & XRT-PC           \\
            &          &              &             &          & UVOT-EV-UVM2     \\
  OV Boo    & \Swift    & 2017-Mar-24  & 00035443009 & 0.4      & XRT-PC     \\
            &          &              &             &          & UVOT-EV-UVM2     \\
  OV Boo    & \Swift    & 2017-Mar-25  & 00035443010 & 0.9      & XRT-PC     \\
            &          &              &             &          & UVOT-EV-UVM2     \\
  OV Boo    & \Swift    & 2017-Mar-28  & 00035443011 & 0.4      & XRT-PC     \\
            &          &              &             &          & UVOT-EV-UVM2     \\
  OV Boo    & \xmmn      & 2017-Aug-03  & 0804620201  & 20.2     & EPIC-FF-THN      \\
            &          &              &             &          & OM-FAST-UVW1     \\
  \hline
  \jt     & \Swift    & 2012-Jul-11  & 00045684001 & 3.0      & XRT-PC           \\
            &          &              &             &          & UVOT-IM-V,B,U,UVW1,UVM2,UVW2     \\
  \jt     & \xmmn      & 2017-May-13  & 0804620101  & 23.8     & XRT-SW-THN       \\
            &          &              &             &          & OM-FAST-UVW1     \\
  \jt     & \Swift    & 2018-Oct-19  & 00045684002 & 1.2      & XRT-PC           \\
            &          &              &             &          & UVOT-IM-V,B,U,UVW1,UVM2,UVW2     \\
           
 \end{tabular}
 \label{tab:obslog}
 
\end{table*}

\subsection{\xmmn}

\xmmn\, observed OV Boo for 20.2\,ks on 2017 August 3 (ObsID 0804620201). The EPIC instruments \citep{StruderEtAl2001, TurnerEtAl2001} were operated in full-frame mode with the thin filter. The optical monitor (OM; \citealt{MasonEtAl2001}) observed the target in fast mode with the UVW1 filter. Effective wavelengths for the OM filters are given in \cite{KirschEtAl2004}. The Reflection Grating Spectrometer (RGS; \citealt{denHerderEtAl2001}) data were taken in spectroscopy mode, but owing to the faintness of the source we have not used these data. This observation was free of high background flaring, except for one short interval near the beginning of the exposure. We obtained 1,565 photons with the EPIC-$pn$ camera.

The same instrumental setups were used for the 23.8\,ks observation of \jt  (2017 May 13, ObsID 0804620101) except that the EPIC instruments were in Small Window mode. This observation was completely unaffected by flaring.

We reduced the data with the \xmmn\, Science Analysis (SAS) software, version 16.1.0. We extracted event lists for the EPIC instruments using the {\tt emproc} and {\tt epproc} scripts, and corrected their timings to the Solar System barycenter  using the {\tt barycen} task. The OM data were reduced using the \texttt{omfchain} script and likewise barycenter corrected.

To extract X-ray spectra, we used circular source extraction regions centered on the target and large circular background regions containing empty sky. The spectra were grouped to 25 counts per bin, and the data from all three EPIC instruments were fitted jointly using version 12.9.1n of Xspec \citep{Arnaud1996}.

\subsection{\Swift}

SDSS1035 was observed by the Neil Gehrels Swift Observatory \citep{BurrowsEtAl2005} on 2012 July 11 (ObsID 00045684001) and 2018 October 19 (ObsID 00045684002). The X-Ray Telescope (XRT) observed in photon counting mode and the Ultraviolet Optical Telescope (UVOT) was operated in imaging mode and obtained images with the V, B, U, UVW1, UVM2, and UVW2 filters in both observations. We did not use data from the Burst Alert Telescope because the source was too faint.

We reduced the X-ray data with the {\tt xrtpipeline} task and corrected the photon arrival times to the Solar System barycenter using the {\tt barycorr} task of Heasoft \citep{Blackburn1995}. The UVOT data were reduced with the standard data reduction tasks described in the UVOT user's guide\footnote{\url{https://swift.gsfc.nasa.gov/analysis/UVOT_swguide_v2_2.pdf}}.

There are  several additional observations taken by \Swift\, of OV Boo. Three were reported previously in \cite{ReisEtAl2013}, but another eight, taken during and shortly after the outburst, have not yet been published. We present an analysis of the previously unpublished observations, which were performed in a variety of observing modes and UVOT filters, summarized in Table \ref{tab:obslog}.

\subsection{\cxo }

The \cxo\ X-ray satellite observed OV Boo serendipitously on 2003 March 14 (ObsID 4071). The second \cxo\ source catalog lists the object with a flux of $(8 \pm 2) \times 10^{-14}$\,\cgs. Since we had a specific spectral model for which we wanted to get a flux from  \cxo, we reduced the data with the \texttt{ciao} software package. As for \XMM\ and \Swift\ we extracted source photons from a circular region containing OV~Boo and background photons from a larger circular source-free region.

\subsection{Optical observations}

\ovbs has almost 160,000 observations performed by the  American Association of Variable Star Observers (AAVSO). We downloaded these data and corrected their timings to the Solar System barycenter using the built-in routines of Astropy, and used data that were obtained close in time to our observations with \xmmn\ to update the eclipse ephemeris.

We note that more archival data are available, namely from the Zwicky Transient Facility (ZTF; \citealt{MasciEtAl2019}) and the Catalina Sky Survey (CSS; \citealt{DrakeEtAl2009}). The ZTF lists 291 measurements in both the $g$ and $r$ filters, and 15 in the $i$ filter obtained between 2018 March to 2019 June, and the CSS has 239 observations in white light obtained between April 2007 and September 2013. These data are not further analysed or used in the current paper, but we show the phase-folded light curves in the Appendix for completeness (Figures A1 and A2). The ZTF data show that \ovbs has completely recovered from the outburst in 2018 and its light curve has re-attained the shape it had at the time of the initial photometric studies by \cite{LittlefairEtAl2007} and \cite{PattersonEtAl2008} with pronounced orbital variability and a bright spot on the accretion disk rim.

\begin{figure*}
\resizebox{\hsize}{!}{
    \includegraphics{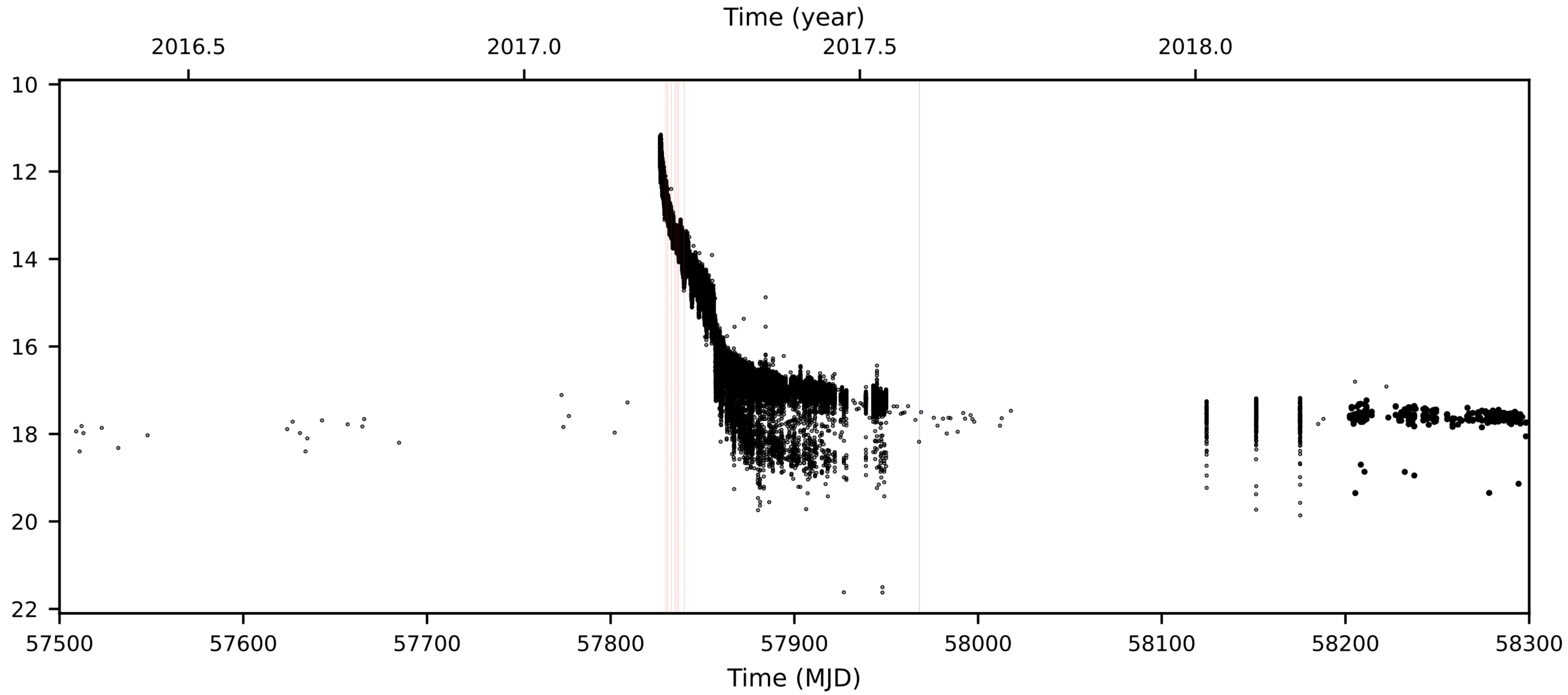}}
    \caption{Long-term optical light curve of \ovbs from 2016 April to 2018 July. AAVSO data in green, visual, or white light extend to about MJD 58200. Thereafter the larger points are ZTF g-band data. The vertical red and blue lines indicate the times of the \Swift\, and \xmmn\, observations, respectively.}
    \label{f:longterm_licu}
\end{figure*}

\section{Analysis and results: \ovb}

\subsection{Orbital light curves and revised eclipse ephemeris}

\begin{figure}
\resizebox{\hsize}{!}{
\includegraphics[angle=270]{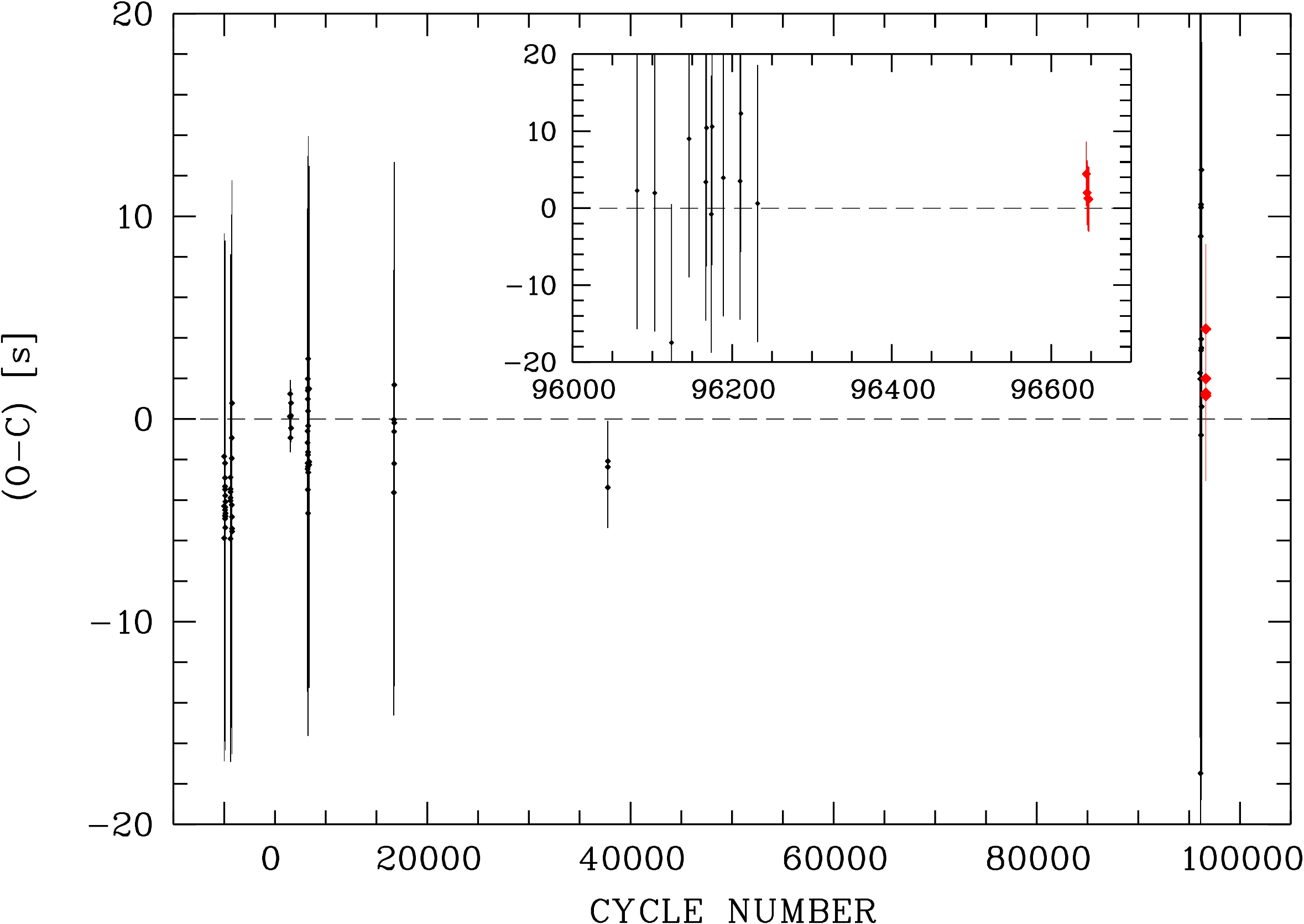}}
    \caption{Observed minus calculated mid-eclipses times for \ovbs with respect to the ephemeris given in Eq.~\ref{e:eph}. The inset shows data obtained between cycles 96000 and 96700 by  AAVSO (black) and with the OM (red).}
    \label{f:ovb_omc}
\end{figure}

In Figure \ref{f:longterm_licu} we show the long-term optical AAVSO and ZTF-green light curve of OV Boo from 2016 April to 2018 July. The times of the X-ray observations are indicated by the vertical lines,  red for \Swift\ just after the peak of the eruption and blue for \xmmn\, in 2017 August after OV Boo had returned to near quiescence brightness. Prior to its eruption in 2017 the object had only been sparsely observed by AAVSO. Many observations were taken at peak and during decay. Vertical features in the AAVSO data, for instance the three vertical lines between MJD 58100 and 58200,  show efforts by the observers to cover the eclipses. The apparent bifurcation in the light curve beginning at about MJD~57870 also shows special attention being paid to the eclipses. Prior to the outburst OV~Boo appeared to have been brightening gradually, increasing by about half a magnitude over several months.

Our new observations performed with \xmmn\ happened about 5 months after the large outburst of the source when the optical brightness had relaxed to a typical bright-phase level of $V\simeq 17.2$, still roughly one mag brighter than during quiescence. All data in this paper were transformed from observed times to orbital phase.

The original OV Boo ephemeris for the eclipse center of \cite{PattersonEtAl2008},
\begin{equation} 
    HJD=2453498.892264(9)+0.0462583411(7) \times E
,\end{equation}
has an accumulated uncertainty of 6\,s at the epoch of our observations with \xmmn\ and needed an update.

To produce a new mid-eclipse ephemeris for OV Boo, we took the existing mid-eclipse timings of \cite{PattersonEtAl2008} and \cite{LittlefairEtAl2007} (we used the timings derived with the $r'$ filter in the latter paper for the documented higher accuracy) and converted them from heliocentric to barycentric timings using Astropy. The first eclipse in the data set marks the epoch for the updated ephemeris.

There were three eclipses in the \hst\ Cosmic Origins Spectrograph (COS) data reported by \cite{UthasEtAl2011}. We downloaded these data, barycenter corrected the timings, and produced a 4\,s binned light curve over the 1425-1900\AA\ wavelength range. We did not mask out any emission lines in these spectra. Then we determined the eclipse ingresses and egresses times using the ``cursor'' method described in \cite{SchwopeThinius2014}, which gives times at half-light between time intervals in and out of the eclipse. The light curve was displayed on a graphics screen and a graphic cursor was used to determine suitable time intervals before, after, and in the eclipse to compute mean brightness levels. The time of half-light at eclipse ingress and egress was then again determined with the cursor guided by reference lines indicating the relevant brightness levels.

Similarly, we used the UV light curves of the OM binned with 2\,s time resolution to determine the midpoints of four eclipses. We did not use timings of the three X-ray eclipses in our \xmmn\ observations because the X-ray eclipse does not seem to trace exactly the same feature as the optical and UV observations do. 

Finally, there are several eclipses in the AAVSO data. We sought events observed well after the end of the 2017 DN outburst and found sixteen, most in 2017 July  and a few in 2019. Eight of these, all from 2017 July, had suitable data quality to measure the ingress and egress. The complete list of eclipses is given in Table \ref{t:ecltimes}. We fitted these with a weighted linear function to derive an ephemeris for mid-eclipse:

\begin{equation}
\label{e:eph}
    \text{BJD(TDB)}=2453498.8930334(34)+0.04625833911(22)\times E
.\end{equation}

\begin{figure}[t]
\resizebox{\hsize}{!}{
\includegraphics{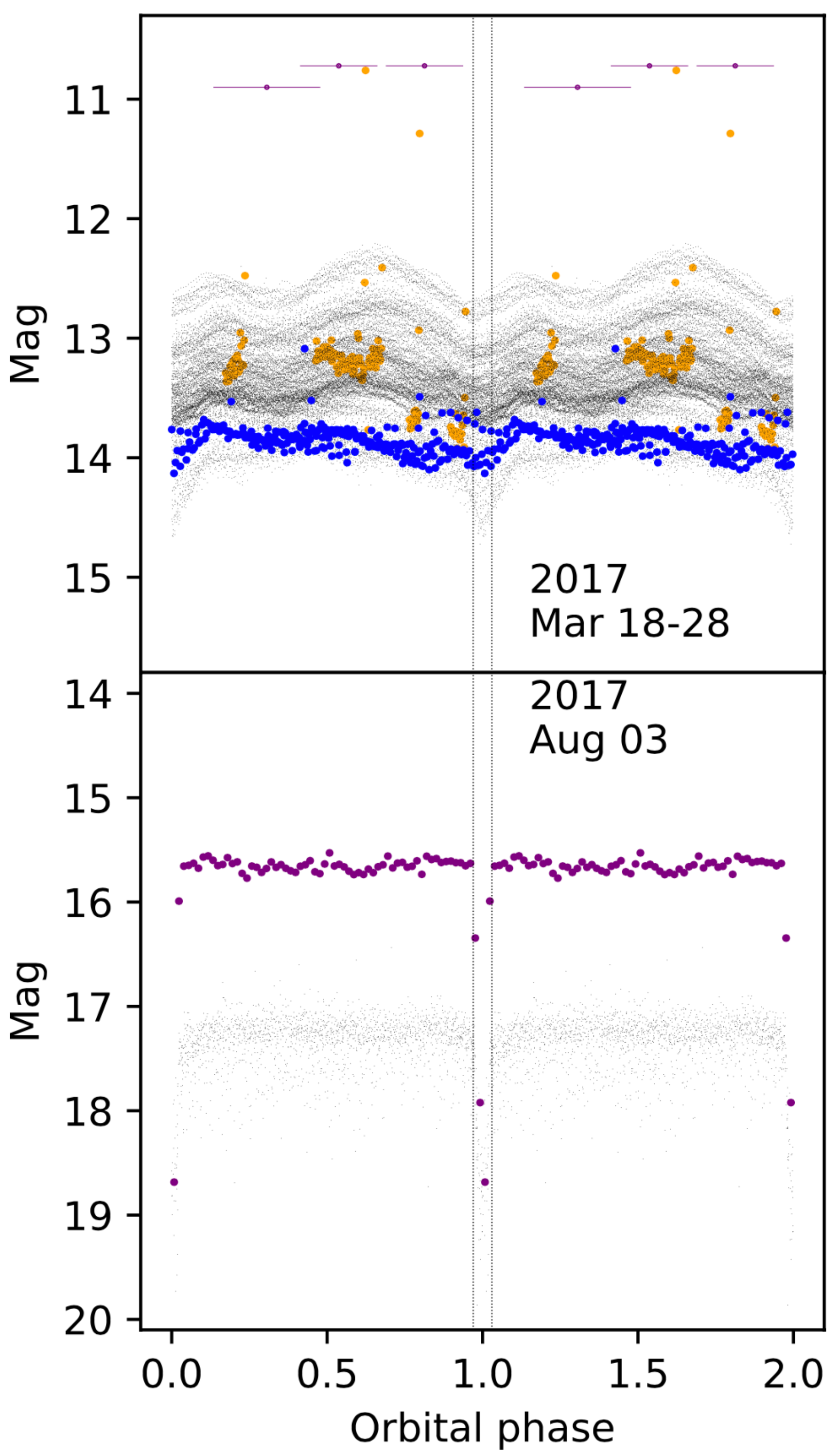}}
\caption{Optical and UV light curves of OV~Boo. \emph{Top panel}: AAVSO visual (black), AAVSO B-band (blue), \Swift\ UVOT M2 (orange), and UVOT W1 (purple). \emph{Bottom panel}: \XMM\ OM W1, binned to 64 $\phi$ intervals (purple) and AAVSO visual (black). OM magnitudes are in the AB system; those from  AAVSO are Vega-mags.}
\label{f:ovboo_optlicus}
\end{figure}

Values in parentheses give the uncertainties in the last digits. The residuals are also listed  in Table~\ref{t:ecltimes} (in the Appendix) and displayed in Fig.~\ref{f:ovb_omc}. The $\chi^2_\nu$ of the fit is only 0.26 (for 80 degrees of freedom), which indicates that the errors were often estimated too conservatively (too big), but we made no attempt to re-adjust the published error bars. We assumed typical error bars for our analysis of half the bin width of the used data. The ephemeris given in Eq.~\ref{e:eph} is used to calculate phases of \ovbs throughout this paper. There is no indication of a significant deviation from linearity in the $(O-C)$ times although the most recent data obtained with the OM in the near UV  and optically by  AAVSO  seem to occur later than predicted by our linear ephemeris. An offset of the eclipse of 2 seconds was accounted for when analyzing and modeling the UV eclipse data (see below).

We also note a slight discrepancy in the eclipse length. For the OM we determine $150.9 \pm 4.2$\,s, which is discrepant at $2.5\sigma$ with the very precise measurement made by \cite{LittlefairEtAl2007} who derived $161.9 \pm 0.2$\,s. This may raise the question whether the pre-outburst optical and the post-outburst UV emission have a common location, which we assume here. However, we cannot exclude some systematic measurement uncertainties here which gives a bias towards an apparently shorter eclipse length. The eclipse times were determined using some reference brightness values in the near vicinity of the eclipse where the brightness was already found to be somewhat reduced with respect to the expectations (see Fig.~\ref{f:ovboo_optlicus}, lower panel). Using a lower reference brightness shifts the ingress to a slightly later time and the egress to an earlier time, hence it affects the ingress and egress times in a similar manner. As a result, the eclipse length might appear too short, but the timing of the eclipse center is expected to be unaffected. 

In Figure \ref{f:ovboo_optlicus} we show the optical and UV behavior  of \ovbs at and around the times of the \Swift\, and \XMM\, observations. In March the object was still fading from its outburst, a phenomenon clearly visible from the AAVSO points. It also appears significantly variable in all wavelengths sampled. By August, OV Boo was behaving very differently. In the optical it had faded to $m_v \sim 17.2$, but, apart from the eclipse, there is very little variability in either the optical or the UV.

\begin{figure}
\resizebox{\hsize}{!}{\includegraphics[clip=]{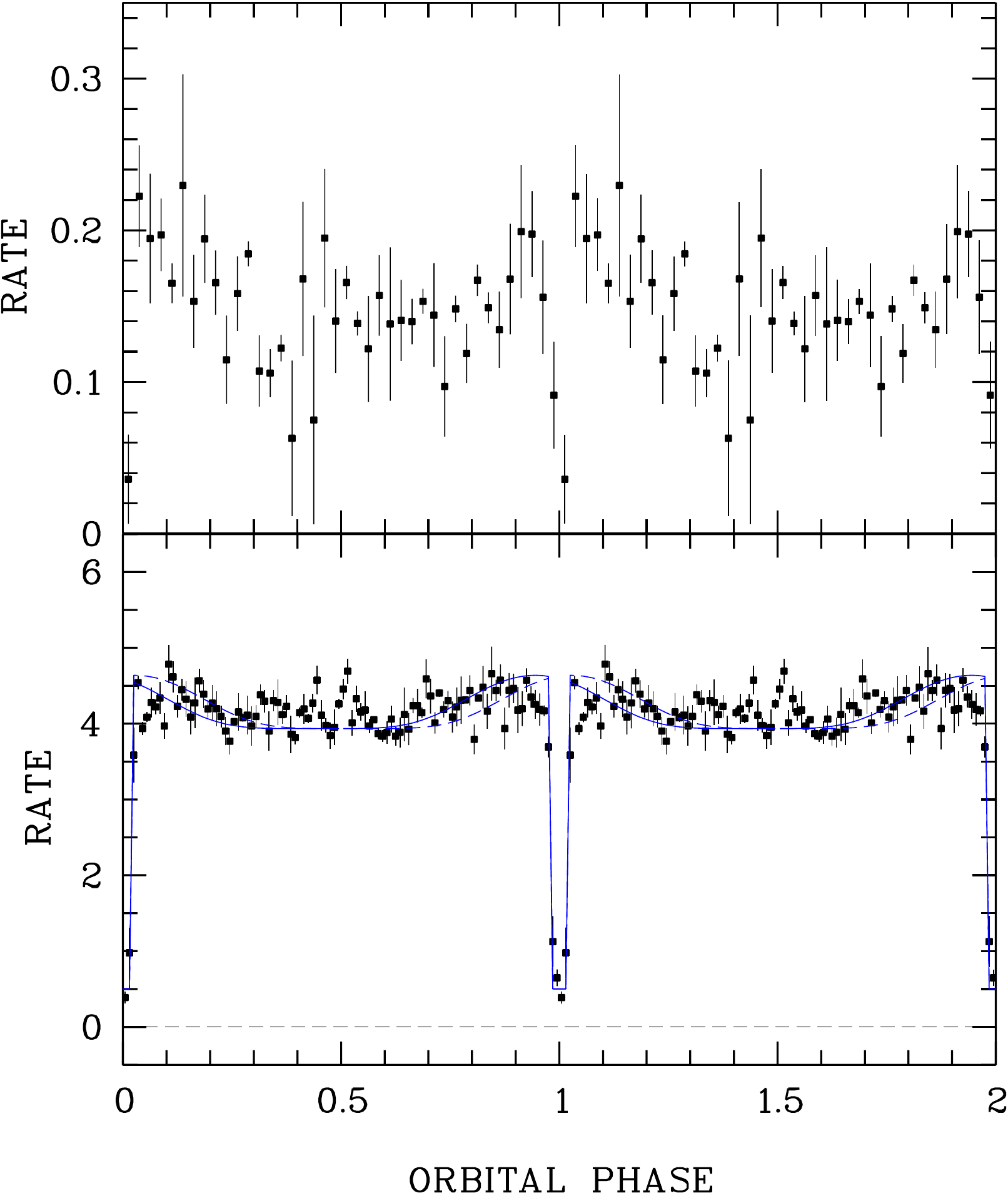}}
\caption{Phase-folded X-ray (top panel) and UV (bottom panel) light curves of \ovbs obtained with \xmmn. Original time-binned data were phase-averaged using phase bins of length 0.025 and 0.01 units, which corresponds to about 100\,s and 40\,s at X-ray and UV wavelengths, respectively. The model curves shown in blue for two different spot locations are explained in the text.}
\label{f:xraypflicu}
\end{figure}

In Figure \ref{f:xraypflicu} we show the phase-folded X-ray and UV light curves of \ovbs from the \XMM\, observation. The data have been grouped into 40 bins, so that the eclipse occupies approximately one bin. The data are very coarse, but the eclipse and a bright phase centered  around the eclipse are plainly visible at both energy ranges. The bright phase is not a very pronounced feature and is difficult to locate precisely in phase.

\begin{figure}[t]
\resizebox{\hsize}{!}{\includegraphics{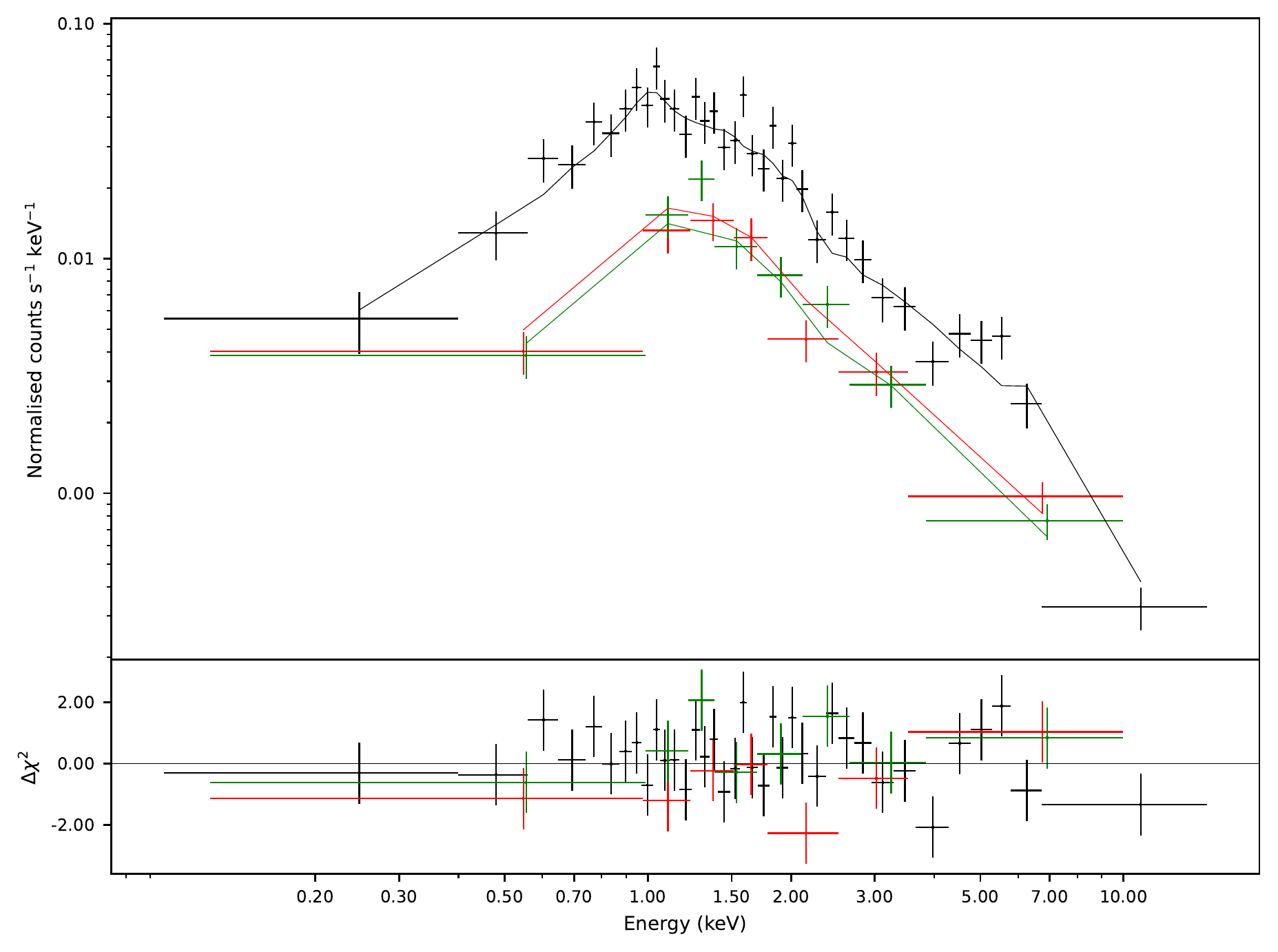}}
   \caption{X-ray spectrum of OV Boo obtained with \xmmn. EPIC$-pn$, MOS1, and MOS2 are plotted in black, red, and green, respectively, with the best fit model overplotted.}
   \label{f:ovboo_xrayspec}
\end{figure}
\begin{table}
   \caption{OV Boo X-ray spectral fit. Fluxes and luminosities are in the 0.2-10\,keV range and uncertainties are $1\sigma$. Fluxes and luminosities are given for the sum of the two plasma emission components.}
   \begin{tabular}{ll}\vspace{1.5mm}
Parameter & Value \\
\hline\\
   nH & $0.24^{+0.09}_{-0.06}\times 10^{22}$\,cm$^{-2}$ \\ \vspace{1.5mm}
   kT & $1.09_{-0.86}^{+5.03}$\,keV \\\vspace{0.8mm}
   norm & $2.0_{-2.0}^{+2.5}\times 10^{-5}$ \\\vspace{0.8mm}
   kT & $9.5_{-3.9}^{+33.5}$\,keV \\\vspace{0.8mm}
   norm & $2.9_{-0.4}^{+0.5}\times 10^{-4}$ \\ \hline
   Abs.   Flux & $5.4\pm 0.2 \times 10^{-13}$\,erg\,s$^{-1}$\,cm$^{-2}$ \\
   Unabs. Flux & $6.9\pm 0.2 \times 10^{-13}$\,erg\,s$^{-1}$\,cm$^{-2}$ \\ 
   Abs. Lum & $2.8\pm 0.2 \times 10^{30}$\,erg\,s$^{-1}$ \\ 
   Unabs. Lum & $3.6\pm 0.4 \times 10^{30}$\,erg\,s$^{-1}$ \\ \hline
   $\chi^2_\nu$ & 1.10 (43) \\
   \end{tabular}
   \label{tab:ovboo_specfit}
\end{table}
\subsection{X-ray spectral analysis of the bright phase}

We fitted the mean \XMM\, X-ray spectrum of OV Boo, excluding the eclipse, with an absorbed (cold interstellar absorption) two-temperature plasma emission model \citep{MeweEtAl1985} plasma (i.e., {\tt tbabs*(mekal+mekal)} in XSPEC), fitting all three instruments jointly between 0.2 and 10.0\,keV. The results are given in Table \ref{tab:ovboo_specfit}. This spectrum is fairly typical of a CV, though at a significantly lower luminosity. The spectrum and residuals are shown in Figure \ref{f:ovboo_xrayspec}.

Although the temperatures of the two plasma components are poorly constrained, we did need to include both. An attempt to fit the spectrum with just one thermal component prevented us from achieving an acceptable spectral fit with $\chi^2_\nu=1.28$ for 45 degrees of freedom. The amount of absorption we find is significantly higher than interstellar. The total column density in the direction of \ovbs is $N_{\rm H} = 1.61 \times 10^{20}$\,cm$^{-2}$ \citep{h14pi}, a factor of 15 lower than that found from the spectral analysis. This hints at intrinsic absorption in the binary.

We then attempted to test the predictions of \cite{PattersonEtAl2008} and \cite{UthasEtAl2011} that OV~Boo is a low-metallicity system by varying the metal abundance of the two-temperature plasma component. We found that the fit   favors a low metallicity, but this parameter also becomes degenerate with the temperature of the hotter MEKAL. In other words, reducing the metallicity caused the temperature to run to an unphysically high value, and both parameters ended up with uncertainties larger than their values: for a metallicity of 0.12 the warm MEKAL's best fit temperature is $30\pm 30$\,keV and for 0.07 it has reached the code hard limit of 79.9\,keV. We therefore conclude that there is weak but inconclusive evidence in the X-ray data that OV~Boo is metal poor, but for the fits we report we have kept the MEKAL values at their default solar metallicity value. 

\begin{figure}[t]
\resizebox{\hsize}{!}{\includegraphics[clip=]{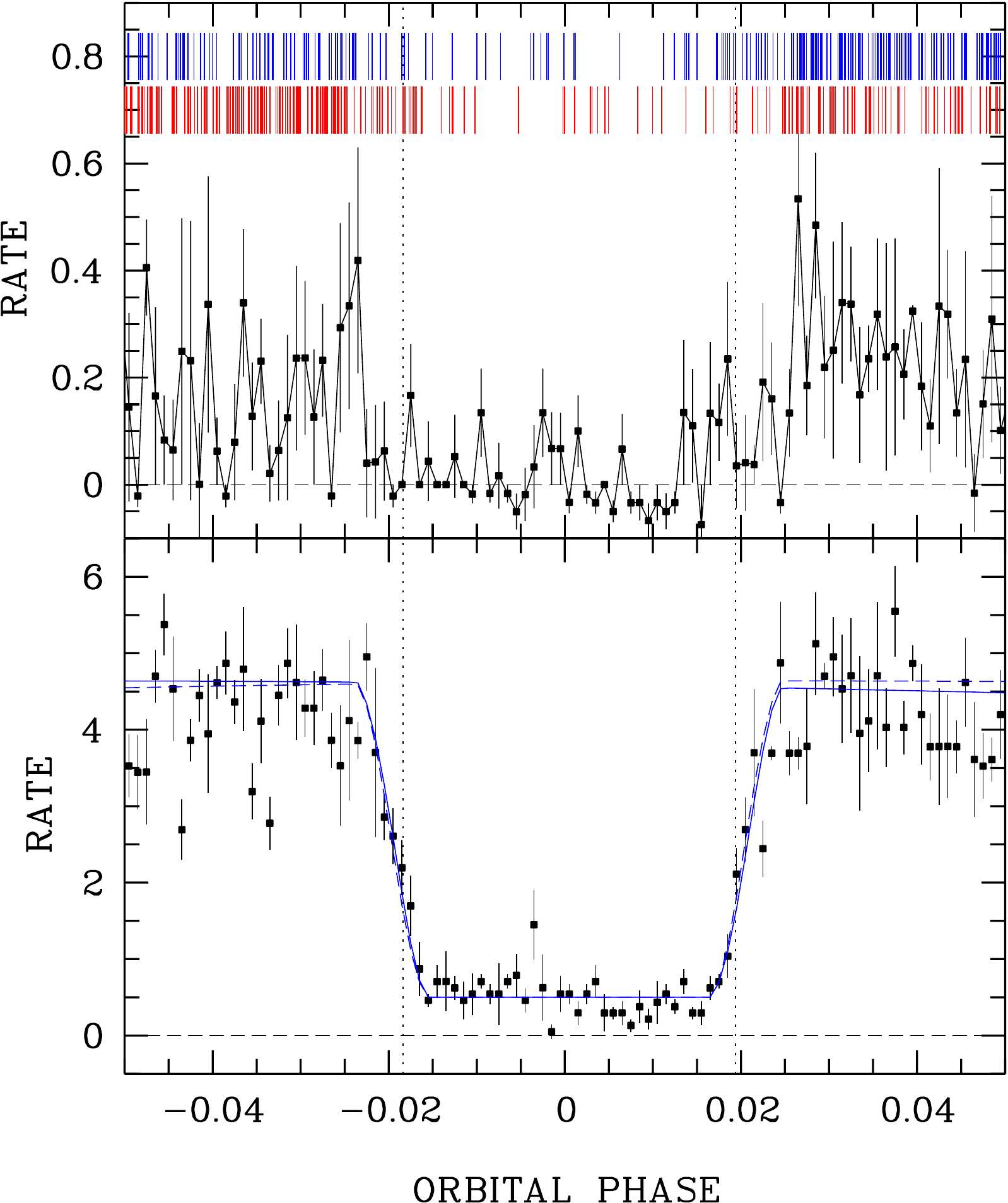}}
\caption{Eclipse profile of OV~Boo at X-ray (\emph{top}) and UV (\emph{bottom}) energies. The original data from EPIC-pn and the OM were binned with 4\,s time resolution and phase-averaged using 1000 phase bins per orbital cycle. The dashed vertical lines indicate the eclipse length of 151\,s measured from the OM data shifted by 2\,s toward a later phase in agreement with the updated ephemeris (see Fig.~\ref{f:ovb_omc}). The vertical blue ticks in the upper panel indicate arrivals of individual photons (pn, MOS1, and MOS2); the red ticks are the same events reflected about $\phi = 0.0$  (corrected for the 2 \,s offset) to highlight asymmetries in the X-ray eclipse. The curves in the lower panel shown in blue are WD plus spot models whose parameters are explained in the text.}
\label{f:ovb_ecl}
\end{figure}

Extrapolating the luminosity measurement to bolometric luminosity is usually a questionable procedure, but since we have little absorption and no significant soft blackbody component, probably safe enough in this case. Using the {\tt cflux} command we found that this model gives an unabsorbed bolometric luminosity of $\sim 5\times 10^{30}$\,erg s$^{-1}$, about 40\% greater than the luminosity in the 0.2-10.0\,keV range reported in Table \ref{tab:ovboo_specfit}. This equates to a mass accretion rate of $\sim 4\times 10^{-13}$\msun\,yr$^{-1}$ using the WD parameters   given above.

\begin{figure}
\resizebox{\hsize}{!}{\includegraphics[angle=270]{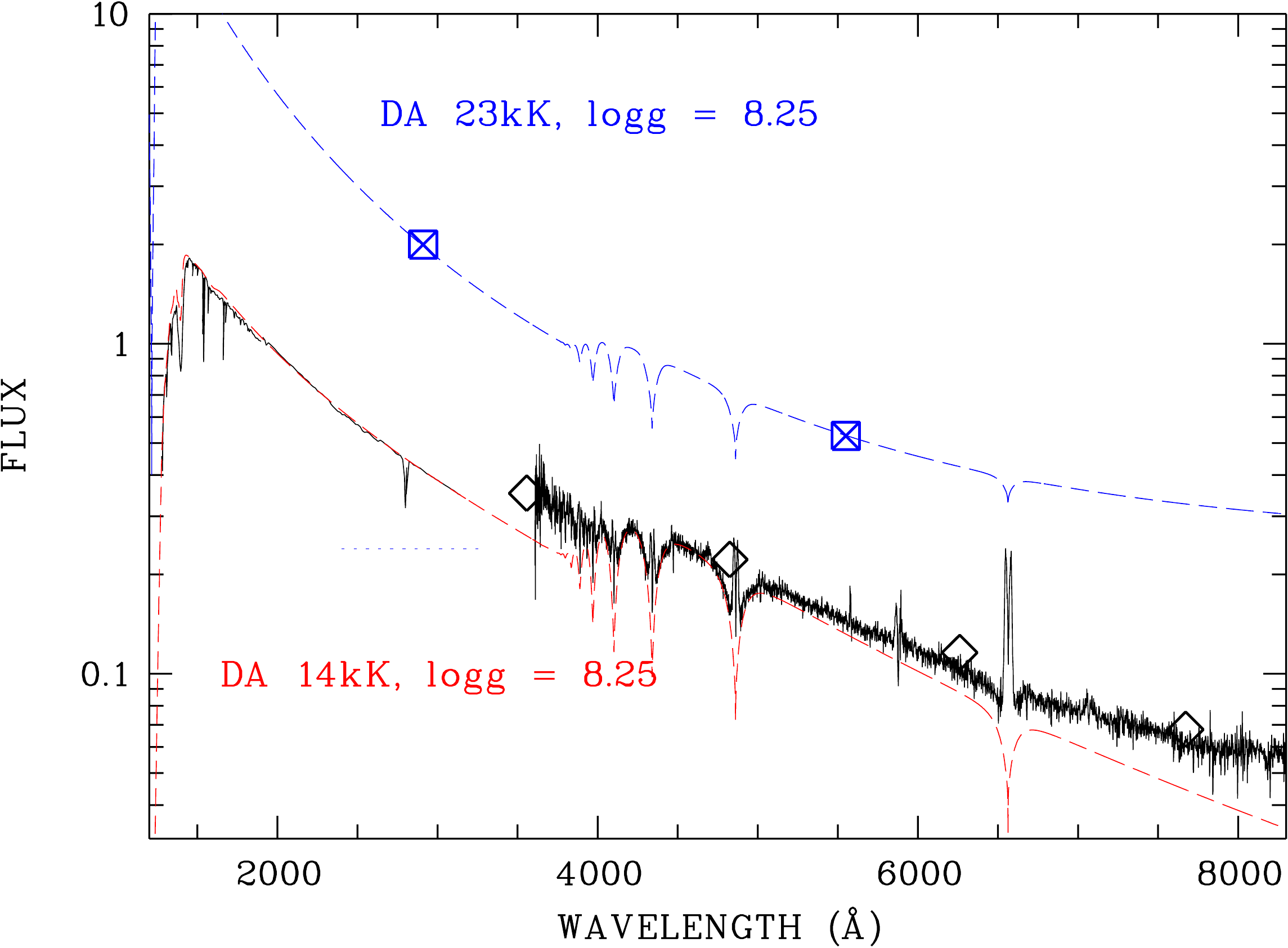}}
    \caption{Optical to UV spectral energy distribution of \ovbs. Shown are the pre-outburst spectra from the {\textit{HST}} \citep{UthasEtAl2011} and the SDSS, and pre-outburst SDSS {\textit{ugri}} photometric data, together with post-outburst photometric data from the OM and  AAVSO (this work). All observed pre-outburst data are shown in black, the two post-outburst data points in blue. White dwarf model spectra for 14 and 23 kK, are shown in red and blue, respectively. Flux units are $10^{-15}$\,\fcgs. The 23kK spectrum was shifted downwards by 0.239 units to account for the non-WD part of the spectrum, whose brightness is indicated by the short blue dashed line at 3000\,\AA.
}
\label{f:ouvsed}
\end{figure}

\subsection{Spot model and optical/UV spectral energy distribution}

In Figure \ref{f:ovb_ecl} we show the eclipse profile of \ovbs at X-ray and UV wavelengths from our observations with \xmmn. The upper panel shows the phase-folded EPIC-pn light curve. 

Apart from the two-second offset of the eclipse center, the UV eclipse profile is symmetric, but it is clear  that the X-ray eclipse leads the formal eclipse midpoint by approximately 0.0035 phase units, or around 15 seconds. To highlight the asymmetry of the X-ray eclipse, the arrival times of X-ray photons from all three EPIC cameras are indicated by vertical lines, both in original time sequence (shown in blue) and mirrored at phase zero (in red), taking the two-second offset into account. We  denote the four contact points of the eclipse of the WD $t_1 \dots t_4$. The X-ray ingress coincides with the first contact point, $t_1$, while the initial egress happens about 10 s before $t_3$. The X-ray egress seems to be complete at $t_4$.

The UV eclipse ingress and egress lasts about $31\pm11$\,s, which is compatible with the size of the WD. At the bottom of the eclipse the flux stays finite at about 10\% of the out-of-eclipse level. There is a gradual decline in the UV flux through the eclipse. We proceed by assuming that all the eclipsed UV flux originates from the WD. The non-eclipsed flux must originate from the accretion disk or structure therein (hot spot). We neglect the variability of the non-eclipsed part, and model the UV data by applying a geometrical model that assumes a WD (and optionally an accretion-heated spot on it) that is gradually obscured and released by the donor star. A grid of pure hydrogen atmosphere models was folded through the response curves of the UV filter. The models used are those of \cite{koester10} for pure hydrogen atmospheres and $\log g = 8.25$. The choice of the gravity has little impact on the continuum brightness which is the only relevant quantity obtained from the spectral models for our light curve model.  A circular spot was placed on the WD and was characterized by its position, its extent, and its temperature distribution. A linear temperature decrease was assumed from a maximum in the spot center to the temperature of the undisturbed WD. The WD surface was divided in small tiles and each tile was assigned a certain brightness according to the chosen model. The binary model then determines which tiles are visible at any given phase. Their summed brightness, modified by foreshortening and limb darkening, then gives a predicted signal as a function of binary phase. 

Since the data appeared rather noisy, several important parameters were pre-defined. The distance was assumed to be $210$\,pc. The parameters of the WD (mass, radius) and the inclination ($=83.6\degr$) were those of  \cite{LittlefairEtAl2007}. The temperature also needs to be pre-defined since the data, shown in Fig.~\ref{f:xraypflicu}, display some flare-like excursions which make it difficult to derive the $T_{\rm WD}$ from fits to the UV light curve. We therefore fixed the temperature of the WD by analyzing the optical--UV spectral energy distribution shown in Fig.~\ref{f:ouvsed}. It shows pre-outburst data from the {\textit{HST} and the SDSS (both the spectrum and the {\textit{ugriz}} photometry) and the post-outburst OM and AAVSO photometry (mean bright phase level). For the presentation in  Fig.~\ref{f:ouvsed} we use Koester's spectra, which are available online \citep{koester10}\footnote{http://svo2.cab.inta-csic.es/theory/newov2/index.php?models=koester2}. The pre-outburst data from the UV to the optical are compatible with the 14,000 K DA WD model spectrum   derived by \cite{UthasEtAl2011}. The model gives an excellent representation of the UV spectra; it underpredicts the observed $u$-band flux from the SDSS and also falls short in the near-infrared. However, such deviations are thought to be  of minor relevance in the current context. The excess emission observed post-outburst at optical (AAVSO) and UV (OM) wavelengths can consistently be attributed to a heated WD of about 23,000\,K (blue symbols and blue curve in Fig.~\ref{f:ouvsed}). A corresponding WD model with uniform temperature of 23,000\,K give an acceptable representation of the observed data. The uncertainty is estimated to be on the  order of 500\,K.

Using a homogeneous WD with a temperature of 23,000\,K also gives an excellent representation of the detailed eclipse shape in the UV (see Fig.~\ref{f:ovb_ecl}). The motivation to also account for a spot with enhanced temperature was derived from enhanced emission around the eclipse. Similar models were used in the past; a graphical illustration of the visibility of such a heated spot is given ~by \cite{gaensicke+98}, among others. Given the low S/N of the data and the high parameter degeneracy, we fixed the spot location and half-opening angle (20\degr) and searched for a spot temperature $T_{\rm spot}$ such that the summed signal of the spot and the undisturbed WD is compatible with the overall light curve of Fig.~\ref{f:xraypflicu}. We found that $T_{\rm spot} = 27,000$\,K gives a satisfactory explanation of the data. The model curves shown in blue in Figs.~\ref{f:xraypflicu} and \ref{f:ovb_ecl} were computed for azimuth values of 20\degr\ (pointing into the quadrant where the accretion stream leaving $L_1$ lies, solid line) and of $-10\degr$ (dashed line), which lies behind the binary meridian. The different parameters cannot be distinguished in the data with high phase resolution centered on the eclipse (Fig.~\ref{f:ovb_ecl}), but only in the total light curve. The model with an azimuth of $-10\degr$
was motivated by the phase of the X-ray hump, whereas the model with $+20\degr$ was motivated by the location of the UV hump.

In summary, the optical to UV spectral energy distribution and the detailed shape of the eclipse observed with the OM can  be nicely explained with a   homogeneously heated WD of 23,000\,K, about 9,000\,K hotter than during quiescence. Some residual UV emission at the bottom of the WD eclipse at a 5\% level must have an origin somewhere away from the WD (i.e.,~in the accretion disk).

\begin{figure}[t]
\resizebox{\hsize}{!}{\includegraphics{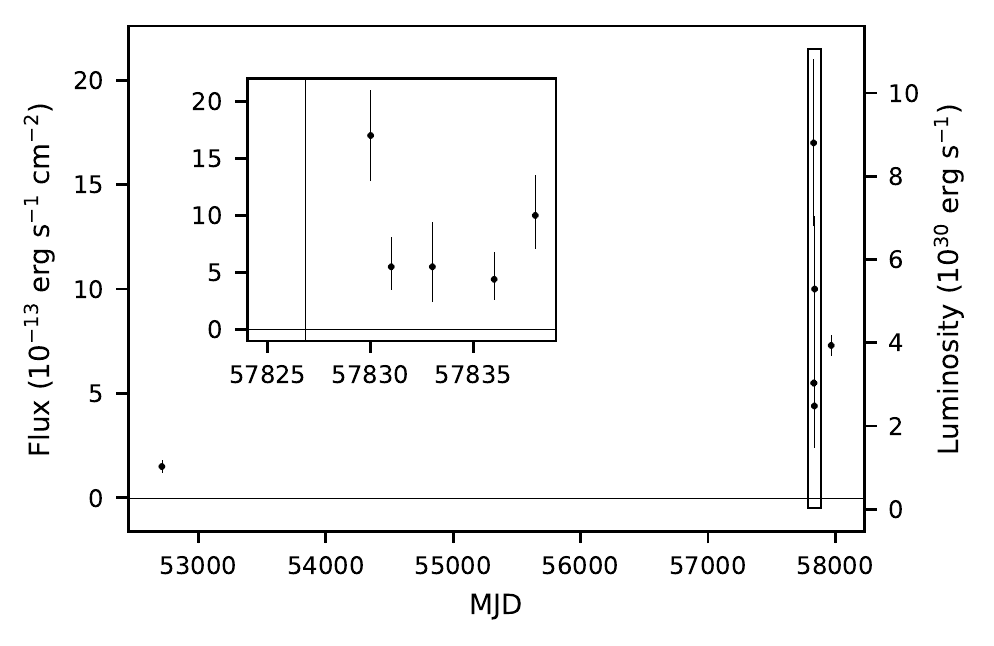}}
    \caption{Long-term X-ray light curve of OV Boo in the 0.2-10\,keV energy range. The \cxo\, and \xmmn\,observations correspond to the first and last points, respectively. The \Swift\, observations of the DN outburst are shown in the inset. Error bars are at $1\sigma$ significance. The luminosity has been calculated using the 210\,pc distance; the relative uncertainty is about 7\%. The vertical line in the inset indicates the time of the first outburst optical observation.}
    \label{f:ovboo_xrayltlicu}
\end{figure}

\subsection{X-ray emission in the eclipse}
We generated an EPIC-pn eclipse image using only photons in the phase interval $(-0.012$:$+0.012)$ and the energy interval $E=0.2 - 5.0$\,keV with a total exposure of 412 seconds. Source detection at a fixed position using the task {\tt edetect\_stack} yields $5.1\pm 3.1$ photons or a rate of $0.0123\pm 0.0075$ at a \emph{DET\_ML} of only 2, consistent with a non-detection. Using the SAS-task {\tt eupper} we determine a $3\sigma$ upper limit count rate in the eclipse of 0.042\,\rat. This is converted into flux by an absorbed plasma emission model with 1\,keV and $N_{\rm H} = 1 \times 10^{20}$\,cm$^{-2}$ and yields $F_{\rm ecl} < 5 \times 10^{-14}$\,\cgs\ or an upper limit luminosity of $L_{\rm ecl} < 3 \times 10^{29}$\,erg\,\rat. The sensitivity is not sufficient to address the question of X-ray emission from the donor star.

\subsection{Long-term X-ray behavior}
The entire long-term X-ray light curve of \ovbs is shown in Figure \ref{f:ovboo_xrayltlicu};  it comprises data obtained with \cxo, \swi, and \xmmn, obtained in quiescence long before the DN outburst, during the outburst, and in the immediate post-outburst phase. We have omitted the three shortest \Swift\ exposures because the source was not detected. The maximum flux observed was about $1.6 \times 10^{-12}$\,\cgs. 

The \cxo\ observation was 4.9\,ks long and delivered 24 photons in a 10" source extraction region, for a background-subtracted source count rate of $4.2^{+1.1}_{-0.9}$\,cts ks$^{-1}$.
There were not enough photons to extract a spectrum, but assuming the same spectral shape as in the \XMM\, observation we infer an absorbed flux of around $1-1.5\times 10^{-13}$\,erg\,s$^{-1}$\,cm$^{-2}$, slightly higher than the catalog flux, but around 5-6 times fainter than in the \XMM\, observation. 

\begin{figure}
 \resizebox{\hsize}{!}{\includegraphics{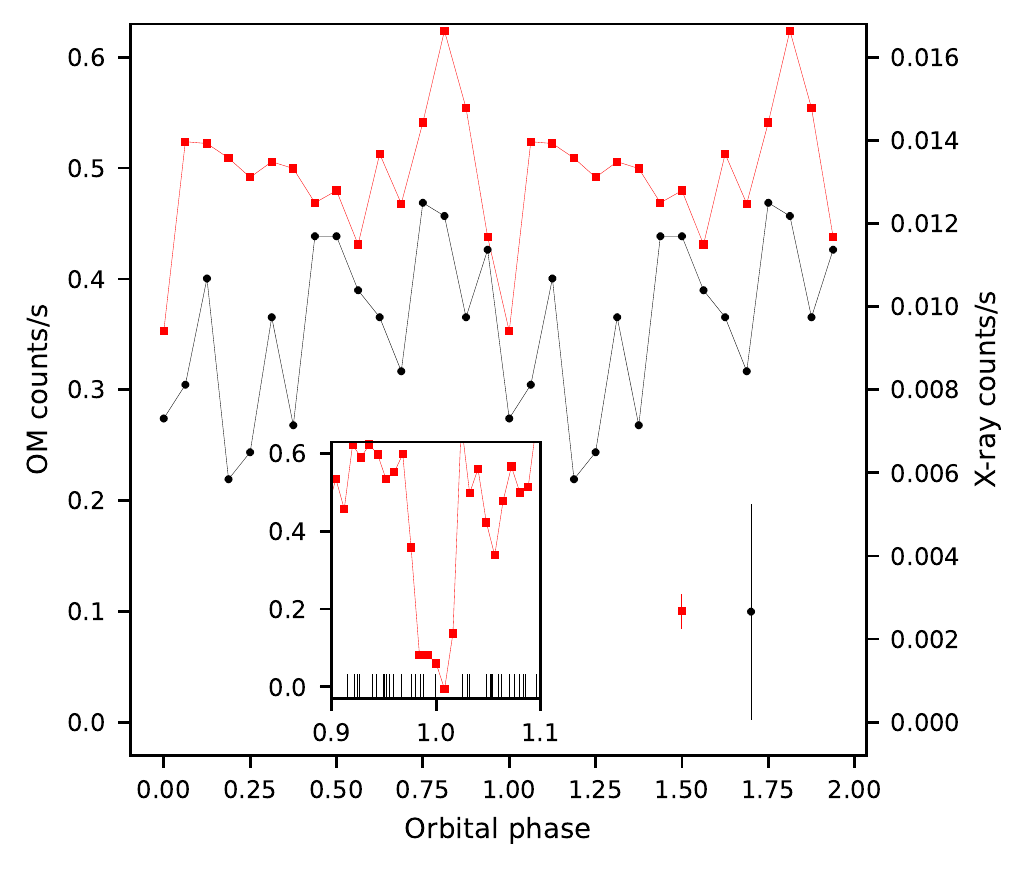}}
    \caption{Phase-folded UV (red squares) and X-ray (black circles) light curve of \jt obtained with \xmmn. The light curve has been duplicated for clarity. Mean $1\sigma$ uncertainties are indicated by the symbols below the curves in the lower right corner. The inset shows the region around the eclipse, with finer binning for the UV points. To avoid cluttering the graph we omit the uncertainties. The vertical black lines indicate the arrival time of the X-ray photons.}
    \label{f:1035pflicu}
\end{figure}

\section{Analysis and results: \jt}

\jt was very faint during the 23.8\,ks \XMM\, observation. Only 215, 20, and 54 photons were detected from the source by the EPIC-$pn$, MOS1, and MOS2 cameras, respectively. Fortunately, the entire observation was unaffected by periods of high background. 
The phase-folded UV and X-ray light curves are shown in Figure \ref{f:1035pflicu}. We used the 82.1\,min period of \cite{LittlefairEtAl2006} and shifted the phase so that the eclipse, which is clearly visible in the UV data, coincides with $\phi=0$. The first eclipse in the UV data occurs at approximately BJD (TDB) = 2,451,521.8636(1).

The eclipses are clearly visible in the phase-folded EPIC$-pn$ data as well, but the data are poor quality and there is a second downward excursion that is probably not real, so we cannot draw a firm conclusion on the existence of an eclipse in the X-ray data.
These data were sufficient to extract coarse spectra for the $pn$ and MOS2 cameras, and produced a single bin for MOS1. These are shown in Figure \ref{f:1035_xrayspec}. We grouped these to a minimum of 16 counts per bin between 0.2 and 10\,keV.

Despite the low spectral resolution we were able to  determine that a single-temperature MEKAL spectrum did not fit the data well: $\chi^2_\nu=1.5$ $(12)$. Formally, this has a chance likelihood of 10\%, but obvious residuals at the low-energy  end suggest a poor model fit. A two-temperature plasma was a much better fit. No absorption was necessary, likely owing to the low counts. The best fit parameters are summarized in Table \ref{tab:1035_specfit}. The spectrum could be described with the combination of a hotter (about 2 keV) collisionally ionized plasma and a cooler component. The cooler component could be described either with a plasma component (parameters listed in Tab.~\ref{tab:1035_specfit}) or with a blackbody, and cannot be statistically discerned. The cool component could be the sought for (or expected) emission from the optically thick boundary layer. Using the observed integrated flux in the 0.2 - 10 keV energy band, the implied instantaneous mass accretion for the 0.94 solar mass WD with $R_{\rm WD} =0.0087$\,\rsun\ is only $8\times10^{-15}$\,\msun yr$^{-1}$ \citep{LittlefairEtAl2006}.

\begin{table}
   \caption{Spectral fit for \jt. Fluxes and luminosities are in the 0.2-10\,keV range, and uncertainties are $1\sigma$.}
   \begin{tabular}{ll}\vspace{1.5mm}
Parameter & Value \\
\hline
kT & $1.7_{-0.3}^{+0.6}$\,keV \\\vspace{0.8mm}
   norm & $6.5_{-1.6}^{+1.8}\times 10^{-6}$ \\\vspace{0.8mm}
   kT & $100^{+50}$\,eV \\\vspace{0.8mm}
   norm & $8.1_{-5.5}^{+11.9}\times 10^{-6}$ \\ \hline
   Flux & $2.0\pm 0.2 \times 10^{-14}$\,erg\,s$^{-1}$\,cm$^{-2}$ \\
   Lum & $1.0\pm 0.2 \times 10^{29}$\,erg\,s$^{-1}$ \\ 
   \hline
   $\chi^2_\nu$ & 0.78 (10) \\
   \end{tabular}
   \label{tab:1035_specfit}
\end{table}

\begin{figure}
 \resizebox{\hsize}{!}{\includegraphics{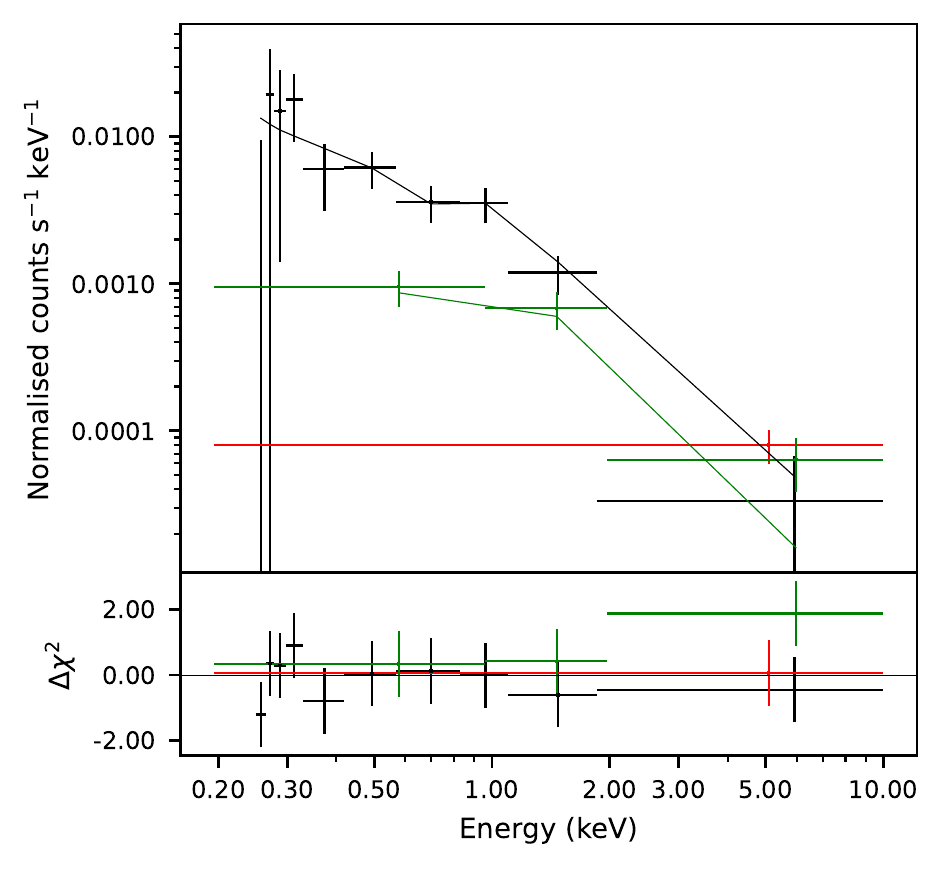}}
    \caption{X-ray spectrum (points with error bars) and fitted two-temperature plasma model (solid lines) of \jt. EPIC$-pn$, MOS1, and MOS2 are plotted in black, red, and green, respectively.}
    \label{f:1035_xrayspec}
\end{figure}

\jt was not detected in X-rays in either of the \Swift\, observations. This is unsurprising; if the source had the same spectral shape and intensity as in the \XMM\, observation, \Swift\, would have detected at most one or two photons in either of its observations.

The optical data from the SDSS and the UV data for this source recorded by \xmmn, \swi, and the Galaxy Evolution Explorer (\GALEX; \citealt{MartinEtAl2005}) are shown in Figure \ref{f:j1035uvmags}. The effective wavelengths of the filters are given in \cite{PageEtAl2014} (for \xmmn), \cite{PooleEtAl2008} (for \swi), and \cite{MorrisseyEtAl2007} (for \GALEX). For the OM we derived two data points, one at the orbital minimum, which represents the maximum contribution of the WD, and one at the orbital maximum, which is due to emission from the hot spot on the accretion disk rim. Part of the variability seen with \swi\ might be due to sampling the light curve at different orbital phases; on the other hand, the source appeared generally around 0.5 magnitudes brighter in 2018 than it was in 2012 in all filters, which probably indicates a real overall brightness increase.

\cite{LittlefairEtAl2008} derive \teff $ = 10100 \pm 200$\,K and a distance of $171 \pm 10$ \,pc for \jt. The revised Gaia distance of 209\,pc requires a parameter update for the WD. Assuming the WD radius (hence mass) is to be fixed by the detailed eclipse model, an increase in \teff\ is required. We found a model with \teff $ = 11250\pm250$\,K that gives the measured WD flux from the deconvolution by \cite{LittlefairEtAl2008}, and that was adequate to reflect the optical--UV spectral energy distribution and to match, not overpredict, the minimum flux observed by \Swift-UVOT. The rather moderate increase in \teff\ nevertheless means a significant increase (about 60\%) in the implied long-term mass accretion rate \citep[from about $1\times 10^{-11}$\,\msun\,yr$^{-1}$ to about $1.6\times 10^{-11}$\,\msun\,yr$^{-1}$, as inferred by using Eq.~2 in][]{townsley_gaensicke09}.

\begin{figure}
 \resizebox{\hsize}{!}{\includegraphics[angle=270,clip=]{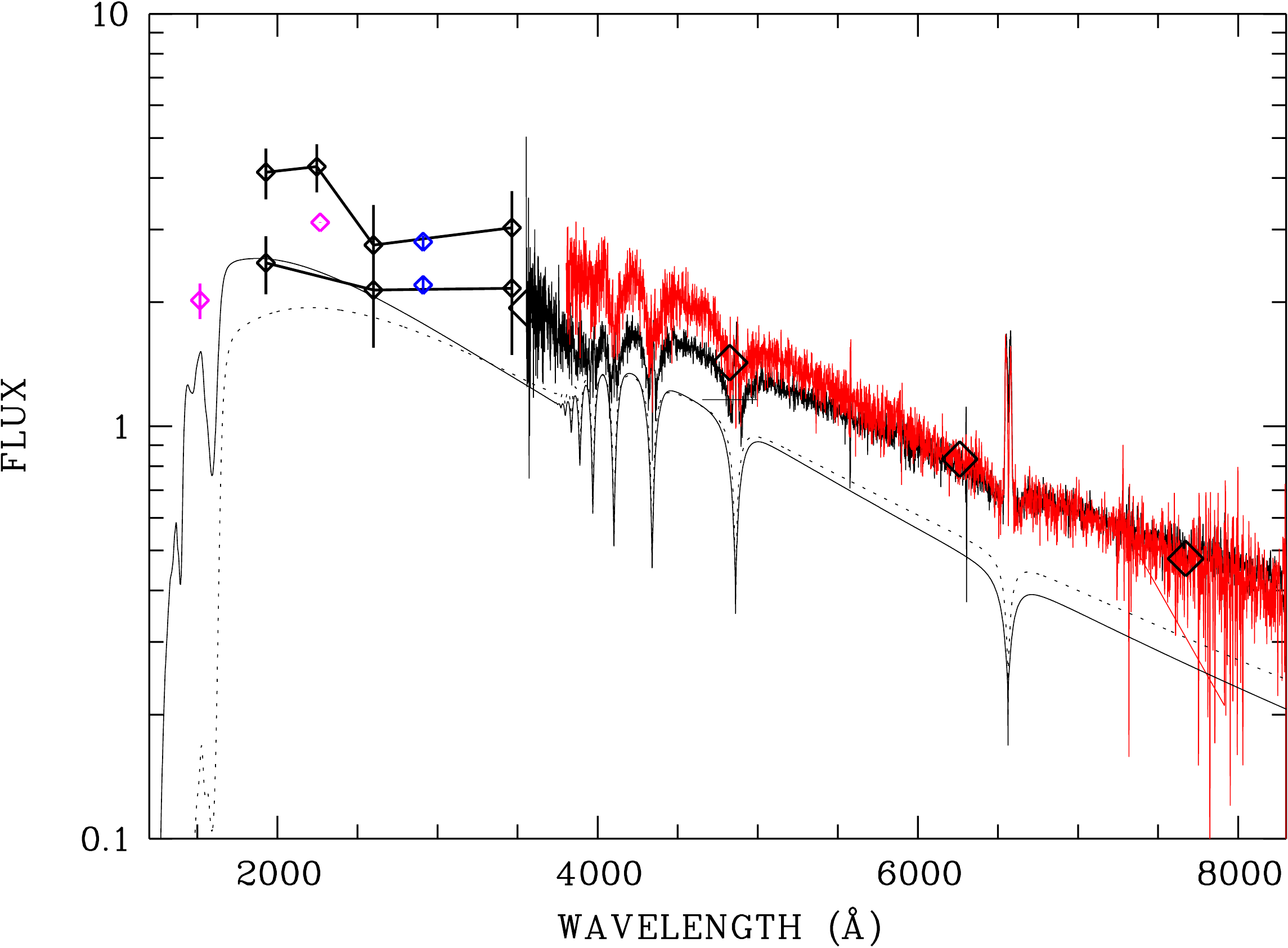}}
   \caption{Optical to UV spectral energy distribution of \jt. Shown are the two spectra obtained by the SDSS (black and red) plus SDSS photometry (black diamonds), \swi-UVOT photometry (black diamonds), \GALEX\ FUV and NUV data (magenta diamonds), and the \xmmn\, OM observations  obtained by us at orbital minimum and maximum (blue diamonds). \swi\ data obtained at the same epoch are connected with lines. The black solid line is a $\log g = 8.25$, $T_{\rm eff} =11500$\,K DA WD model spectrum scaled to a distance of 209 pc. Flux units are $10^{-16}$\,\fcgs.}
   \label{f:j1035uvmags}
\end{figure}


\section{Discussion}
We  analyzed X-ray, ultraviolet, and optical observations of two eclipsing period-bouncing CVs (i.e.,~CVs with a degenerate donor star): \ovbs and \jt. Both objects were discovered at X-ray wavelengths indicating ongoing accretion. \ovbs shows an X-ray eclipse, which hints at an emission region close to the WD, likely the boundary layer. The situation is less clear in \jt due to its weaker X-ray signal.

Although the two stars are at about the same distance, \ovbs was much brighter at X-ray and UV wavelengths thanks to a higher instantaneous accretion rate and a recent (rare) DN outburst. The DN outburst happened 142 days before our observations with \xmmn\ were scheduled. These data gave deeper insight into the origin of the X-ray and UV emission through a detailed eclipse analysis than those of \jt, and are discussed first in the following paragraphs.

The main findings for \ovbs can be summarized as follows. The X-ray and UV light curves outside the eclipse are not completely flat but seem to show some flickering behavior and some excess emission centered around phase zero. Apart from the flickering, the detailed UV eclipse light curve and the mean out-of-eclipse  brightness are compatible with a WD with \teff $= 23$\,kK, some 9000\,K hotter than found before the DN outburst. Both the UV and the optical magnitude found contemporaneously are compatible with this temperature. Approximately 10\% of the UV flux does not go into eclipse. This remaining flux displays some decrease through the eclipse. It appears possible that this remaining flux originates from the hot spot where the accretion stream hits the outer rim of the accretion disk. The detailed eclipse models by \cite{LittlefairEtAl2007} suggest that the total eclipse of the hot spot   begins at phase 0.1, which does not seem to comply with our data. We do not want to overinterpret our data, but we cannot exclude some extra emission at this phase from another location in the binary.

The X-ray eclipse seemingly has a faster ingress and egress than the UV eclipse. In addition, the eclipse center seems to be shifted in phase to occur earlier by 0.0035 phase units or about 15 s. While the eclipse ingress seems to be sharp, the egress may be structured and happen in two steps.  The mean X-ray spectrum is highly absorbed, a factor of 15 higher than the galactic column density in the direction of the star, and is purely thermal. Taking these findings together a picture emerges in which all the observed X-ray emission is  located in the boundary layer (BL) between the inner rim of the disk and the WD surface \citep[see the thorough discussion of quiescence DN X-ray emission in ][]{mukai17}. The high column density suggests that any potential optically thick emission from the BL is absorbed and not accessible to us. Some extra absorption by matter (e.g., at the hot spot) might be responsible for the early ingress. This matter is no longer in the line of sight at eclipse egress so that this feature is observed at the expected time and/or phase.

The implied instantaneous mass accretion rate from the \XMM\, observations is low, $\sim 4\times 10^{-13}$\msun\,yr$^{-1}$, but still a factor of 5-6 higher than during a serendipitous \cxo\ observation taken a few years before the outburst. Whether the mass transfer rate from the donor was higher than normal or the mass flow rate through the disk still enhanced in the post-outburst phase cannot be determined from our data.

We could derive an upper limit for the X-ray emission in the eclipse of $L_{\rm ecl} < 3 \times 10^{29}$\,erg\,\rat. We are not reaching sufficient sensitivity to constrain X-ray emission from the donor star. 

\jt is much fainter than OV Boo, and therefore difficult to study. It is less luminous by a factor of $\sim 40$ in both X-rays and UV. The instantaneous accretion rate of \jt is lower than that of \ovbs by around 1.5 orders of magnitude (a factor of about 50),  $8\times 10^{-15}$\,\msun\,\yrat. This rate is even lower than that of the only magnetic period bouncer studied by \cite{stelzer+17} who found a mass accretion rate of $3.2 \times 10^{-14}$\,\msun\,\yrat. 

The two objects have the same inclination of about 83\degr. \ovbs shows a high level of intrinsic absorption, whereas \jt does not show any sign of absorption in its X-ray spectrum, indicating rather different shapes of the accretion disk and hot spots on its rim. Instead of absorption \jt shows  some sign of the expected soft emission from the boundary layer, but its faintness does not allow us to derive conclusive evidence on its existence. 

Through the inclusion of the UV in our analysis we were able to derive revised temperatures for the WDs in both objects. The high temperature found in our observation in \ovbs is due to the recent DN outburst, the long-term average temperature of the WD is  better described with the value of  14,000\,K  found by \cite{UthasEtAl2011}. The larger distance to J1035 together with the various photometric data obtained in the UV led to the revised \teff\ of 11,250\,K. The updated WD temperatures together with their precisely determined masses can be used to infer the long-term mass accretion rates for both objects. Using Eq.~2 of \cite{townsley_gaensicke09} the rates are $\log{\dot{M}} = -10.33$\,\msun\,\yrat\ and $-10.79$\,\msun\,\yrat\ for \ovbs and J1035, respectively. These values place the two objects outside the tracks of \cite{willems_kolb05} that were used by \cite{LittlefairEtAl2008} to discuss the evolutionary status of these short-period objects. Whether this deviation is significant for the class needs to be seen through an enlarged sample of objects.

\section{Conclusion}
This last section is   devoted to a quick discussion of the prospect of enlarging the sample through identification programs of the X-ray surveys with eROSITA.
eROSITA has just completed its first X-ray all-sky survey \citep{predehl+20}, eRASS1. The first survey reached a point-source limiting flux in the soft band (0.2-2.3 keV) of $5\times 10^{-14}$\,\cgs, the final depths after completion of the planned eight independent surveys in the ecliptic equatorial regions will be $1.1 \times 10^{-14}$\,\cgs. If we  assume that the flux of J1035 is   representative of the period bouncing, low-luminosity CVs at 210 pc, then the radii to search for more objects of this kind as X-ray emitters in the first and the full eROSITA all sky survey are 133 and 280\,pc, respectively. How many bouncers can we detect in the corresponding volumes? The volume surveyed in eRASS1 corresponds to the 150 pc sample surveyed by \cite{pala+20} who only have three candidate period bouncers in their sample of 42 objects (with a possible missing 12 objects of any CV subtype for 100\% completeness). Predicted space densities of \cite{belloni+20} (their model 324509 in Table 1) for period bouncers are as high as $\rho_0 = 17 \times 10^{-6}$\,cm$^{-3}$ with a factor of two uncertainty. In agreement with this, \cite{hernandez-santisteban+18} derive a 2$\sigma$ upper limit for period bouncing objects they call dead CVs of $\rho_0 = 2 \times 10^{-5}$\,cm$^{-3}$. Using a constant density, justified by the expected large scale-height of more than 400 pc for this old population, and the given upper limit we expect a maximum of 24 objects in a volume of radius 150 pc and about 180 in a volume with radius 280\,pc, to be reached after finishing the eROSITA all-sky surveys.

For the two objects studied here we determined the flux ratio between the optical and the X-ray flux in quiescence to check the feasibility of eROSITA follow-up. For \ovbs we use $g=18.31$, $f_{\rm X} = 1 \times 10^{-13}$\,\cgs\ (\cxo), for J1035 we use $g=18.79$ and $f_{\rm X} = 2 \times 10^{-14}$\,\cgs\ to obtain flux ratios $\log(f_{\rm X}/f_{\rm opt}) = \log{f_{\rm X}} - 0.4 \mathrm{mag} + 5.37$ of $-0.3$ and $-0.8$, respectively. These numbers are close to those found for ordinary CVs, and should help to separate accreting objects from coronal emitters. Very late-type stars at the bottom of the main sequence may reach similarly high values, but will be distinguished from the low-luminosity CVs through their distinct red color as opposed to the very blue color of the period bouncers \citep[see][their Fig.~11]{comparat+20}. At the given flux limit all candidates will have counterparts in the Gaia catalogs. Hence, the feasibility to build statistically meaningful samples through comprehensive X-ray identification programs is excellent. 

\begin{acknowledgements}
We thank an anonymous referee for carefully reading the original manuscript and giving useful advice. 

This work was supported by the German DLR under contracts 50 OR 1405, 50 OR 1711, 50 OR 1814, and 50 OX 1901. We acknowledge with thanks the variable and comparison star observations from the AAVSO International Database contributed by observers worldwide and used in this research. Many of the AAVSO observations were the result of Centre for Backyard Astronomy (CBA) campaigns. We thank the AAVSO observers who made available their data via the AAVSO and used here to refine the orbital period: David Cejudo Fernandez and Lew Cook from Spain and the US. This research has made use of the APASS database, located at the AAVSO web site. Funding for APASS has been provided by the Robert Martin Ayers Sciences Fund. This work has made use of Astropy \citep{Astropy2013,Astropy2018}. This research has made use of data, software and/or web tools obtained from the High Energy Astrophysics Science Archive Research Center (HEASARC), a service of the Astrophysics Science Division at NASA/GSFC and of the Smithsonian Astrophysical Observatory's High Energy Astrophysics Division.

\end{acknowledgements}

\bibliographystyle{aa}
\bibliography{bibli}


\begin{appendix}

\section{Eclipse times used to derive the ephemeris of \ovbs}

\begin{table*}[ht]
 \caption{Eclipse midpoint timings for OV Boo. Listed are the cycle number, the times of eclipse center in Barycentric Julian Days in the TT time system (terrestrial time), the uncertainty of the timing, the deviation from the linear ephemeris given in Eq.~\ref{e:eph}, and an observer code. Observer codes P, L, H, A, and OM stand for \cite{PattersonEtAl2008}, \cite{LittlefairEtAl2007}, \hst, the AAVSO, and the Optical Monitor on \xmmn, respectively.  }
 \begin{tabular}{rlrrr|rlrrr}
Cycle & BJD(TDB) & Unc. (s) & O-C (s) & Obs. & Cycle & BJD(TDB) & Unc. (s) & O-C (s) & Obs.\\
\hline
    0 & 2453498.89301 &11.0  &    -1.9  &  P &     8234 & 2453879.78421 &11.0  &     1.4  &  P \\              
     1 & 2453498.93924 &11.0  &    -4.3  &  P &    8236 & 2453879.87667 &11.0  &    -3.5  &  P \\              
     2 & 2453498.98548 &11.0  &    -5.9  &  P &    8237 & 2453879.92298 &11.0  &     1.0  &  P \\              
    82 & 2453502.68616 &11.0  &    -4.8  &  P &    8238 & 2453879.96925 &11.0  &     2.0  &  P \\              
    83 & 2453502.73244 &11.0  &    -2.9  &  P &    8254 & 2453880.70934 &11.0  &    -1.8  &  P \\              
    84 & 2453502.77868 &11.0  &    -4.5  &  P &    8255 & 2453880.75560 &11.0  &    -1.6  &  P \\              
    85 & 2453502.82495 &11.0  &    -3.5  &  P &    8256 & 2453880.80185 &11.0  &    -2.4  &  P \\              
    86 & 2453502.87121 &11.0  &    -3.3  &  P &    8257 & 2453880.84814 &11.0  &     0.4  &  P \\              
    87 & 2453502.91745 &11.0  &    -4.9  &  P &    8258 & 2453880.89434 &11.0  &    -4.7  &  P \\              
    88 & 2453502.96374 &11.0  &    -2.2  &  P &    8275 & 2453881.68082 &11.0  &     3.0  &  P \\              
   107 & 2453503.84263 &1.1.  &    -3.8  &  P &    8276 & 2453881.72704 &11.0  &    -0.3  &  P \\              
   108 & 2453503.88887 &11.0  &    -5.4  &  P &    8277 & 2453881.77332 &11.0  &     1.5  &  P \\              
   109 & 2453503.93514 &11.0  &    -4.4  &  P &    8278 & 2453881.81953 &11.0  &    -2.6  &  P \\              
   129 & 2453504.86031 &11.0  &    -4.1  &  P &    8365 & 2453885.84401 &11.0  &    -2.3  &  P \\              
   130 & 2453504.90656 &11.0  &    -4.8  &  P &    8366 & 2453885.89027 &11.0  &    -2.1  &  P \\              
   131 & 2453504.95282 &11.0  &    -4.7  &  P &    8367 & 2453885.93657 &11.0  &     1.5  &  P \\              
   622 & 2453527.66565 &11.0  &    -5.9  &  P &   16710 & 2454271.86984 &11.0  &    -3.6  &  P \\              
   623 & 2453527.71193 &11.0  &    -4.0  &  P &   16711 & 2454271.91614 &11.0  &    -0.0  &  P \\              
   624 & 2453527.75819 &11.0  &    -3.9  &  P &   16731 & 2454272.84120 &11.0  &    -0.6  &  P \\              
   625 & 2453527.80446 &11.0  &    -2.9  &  P &   16732 & 2454272.88754 &11.0  &    -2.2  &  P \\              
   626 & 2453527.85071 &11.0  &    -3.6  &  P &   16752 & 2454273.81273 &11.0  &    -0.2  &  P \\              
   627 & 2453527.89697 &11.0  &    -3.5  &  P &   16753 & 2454273.85901 &11.0  &     1.7  &  P \\              
   752 & 2453533.67928 &11.0  &    -1.9  &  P &   37774 & 2455246.25550 &2.0   &   -3.4   & H  \\              
   753 & 2453533.72555 &11.0  &    -0.9  &  P &   37775 & 2455246.30177 &2.0   &   -2.4   & H  \\              
   754 & 2453533.77177 &11.0  &    -4.2  &  P &   37777 & 2455246.39429 &2.0   &   -2.1   & H  \\              
   774 & 2453534.69693 &11.0  &    -4.8  &  P &   96081 & 2457943.44054 &18.0   &    2.3   & A \\               
   775 & 2453534.74318 &11.0  &    -5.6  &  P &   96103 & 2457944.45822 &18.0   &    2.0   & A \\               
   776 & 2453534.78944 &11.0  &    -5.4  &  P &   96124 & 2457945.42942 &18.0   &  -17.5   & A \\               
   777 & 2453534.83577 &11.0  &     0.8  &  P &   96146 & 2457946.44741 &18.0   &    9.0   & A \\               
  6482 & 2453798.739589& 0.7  &     0.1  &  L &   96167 & 2457947.41877 &18.0   &    3.4   & A \\               
  6500 & 2453799.572252& 0.7  &     1.2  &  L &   96168 & 2457947.46511 &18.0   &   10.4   & A \\               
  6502 & 2453799.664756& 0.7  &     0.1  &  L &   96174 & 2457947.74253 &18.0   &   -0.8   & A \\               
  6521 & 2453800.543664& 0.7  &     0.1  &  L &   96175 & 2457947.78892 &18.0   &   10.6   & A \\               
  6524 & 2453800.682427& 0.7  &    -0.9  &  L &   96189 & 2457948.43646 &18.0   &    3.9   & A \\               
  6569 & 2453802.764065& 0.7  &     0.2  &  L &   96210 & 2457949.40788 &18.0   &    3.5   & A \\               
  6586 & 2453803.550464& 0.7  &     0.8  &  L &   96211 & 2457949.45424 &18.0   &   12.3   & A \\               
  6587 & 2453803.596708& 0.7  &    -0.5  &  L &   96232 & 2457950.42553 &18.0   &    0.6   & A \\               
  8213 & 2453878.81274 &11.0  &    -2.5  &  P &   96644 & 2457969.48401 &4.2   &    4.4   & OM  \\             
  8214 & 2453878.85902 &11.0  &    -0.6  &  P &   96645 & 2457969.53024 &4.2   &    2.0   & OM  \\             
  8215 & 2453878.90526 &11.0  &    -2.2  &  P &   96646 & 2457969.57649 &4.2   &    1.3   & OM  \\             
  8216 & 2453878.95153 &11.0  &    -1.2  &  P &   96647 & 2457969.62275 &4.2   &    1.2   & OM  \\             
 \end{tabular}
 \label{t:ecltimes}
\end{table*}

\begin{figure}
\resizebox{\hsize}{!}{
    \includegraphics{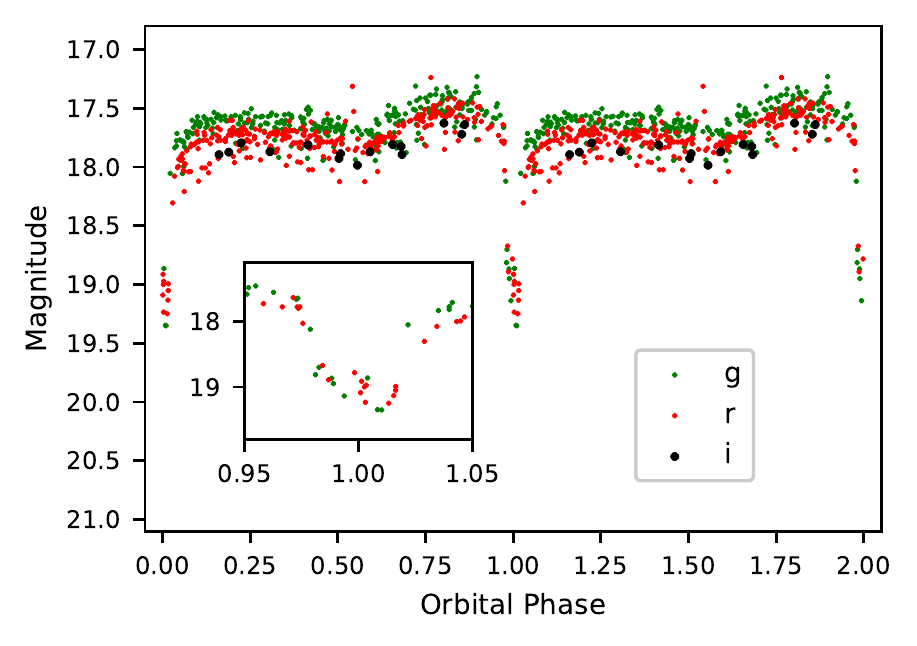}}
   \caption{Phase-folded light curve of OV~Boo from the ZTF (2018 March - 2019 June). The light curve has been duplicated for clarity. The inset shows the $r$ and $g$ data in the region of the eclipses.}
    \label{f:ovb_ztf}
\end{figure}

\begin{figure}
\resizebox{\hsize}{!}{
    \includegraphics{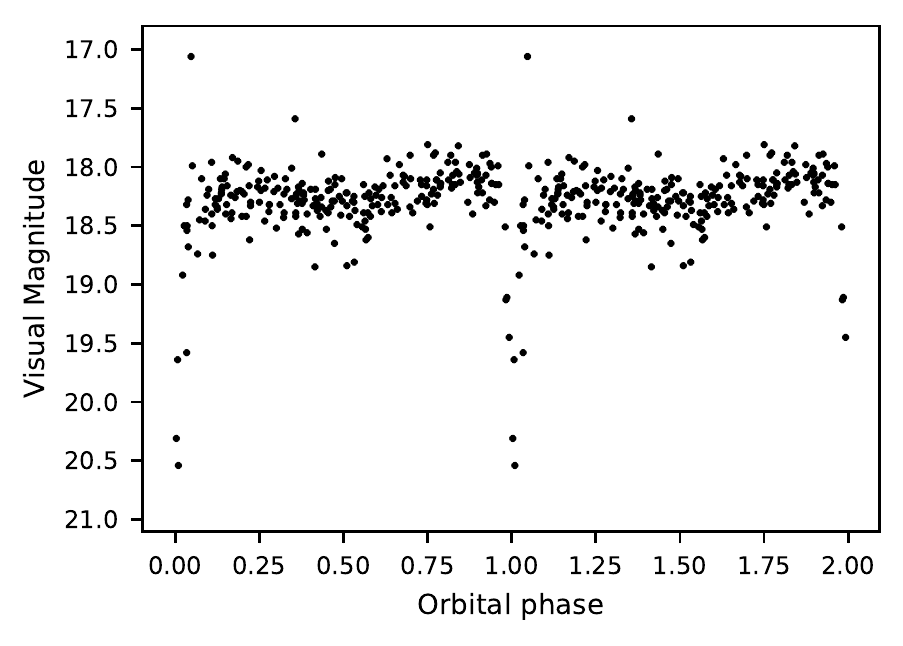}}
    \caption{Phase-folded light curve of OV~Boo from CSS (2007-2013). The light curve has been duplicated for clarity.}
    \label{f:ovb_css}
\end{figure}

\end{appendix}
\end{document}